\documentclass[10pt,twocolumn,journal]{IEEEtran}
\usepackage{array}
\usepackage{multirow}
\usepackage{multicol}
\usepackage{tablefootnote}
\usepackage{makecell}
\usepackage[export]{adjustbox}
\usepackage{color}
\usepackage{longtable}

\usepackage[numbers]{natbib}  

\usepackage[utf8]{inputenc}
\usepackage{authblk}  

\usepackage[utf8]{inputenc}
\usepackage[T1]{fontenc}
\usepackage{lmodern}

\usepackage{amsmath, amssymb}

\usepackage{graphicx}       
\usepackage{caption}        
\usepackage{subcaption}     
\usepackage{float}          
\usepackage{placeins}       

\usepackage{hyperref}
\usepackage[capitalise]{cleveref}  

\usepackage{xspace}         

\title{Revealing a transitional epoch of large-scale cosmic anisotropy in the quasar distribution}

\author[1]{Amit Mondal \thanks{E-mail: amitmondal.bwn95@gmail.com,}}
\author[1]{Biswajit Pandey \thanks{biswap@visva-bharati.ac.in (Corresponding author),}}
\author[1]{Krishna Ghosh \thanks{krishnaghosh879@gmail.com}}

\affil[1]{Department of Physics, Visva-Bharati University, Santiniketan, 731235, India}

\date{}  

\begin{document}

\maketitle

\begin{abstract}
The Cosmological Principle posits that the Universe is isotropic on the largest scales. While widely supported, this foundational assumption remains testable. We analyse the angular distribution of over one million quasars from the Gaia-unWISE catalogue using Renyi entropy, a multiscale statistical measure sensitive to higher-order clustering. Dividing the sample into three redshift bins, we find that both the low- and high-redshift distributions are statistically consistent with isotropy. However, at intermediate redshift ($1 \leq z < 2.2$), we detect a statistically significant and scale-dependent anisotropy that persists under stringent masking, suggesting a physical origin. We interpret this as evidence for a transitional epoch in cosmic history, during which large-scale structures such as superclusters became prominent before their growth was gradually damped by the onset of accelerated expansion. These findings position Renyi entropy as a powerful probe of cosmic evolution and highlight the potential thermodynamic links between structure formation, entropy dissipation, and the emergence of large-scale isotropy.
\end{abstract}
\vspace{1em}

\section*{Cosmic isotropy and quasar sky}
One of the foundational assumptions of modern cosmology is the Cosmological Principle, the idea that the Universe is statistically homogeneous and isotropic on sufficiently large scales. This principle underpins the Friedmann-Lemaitre-Robertson-Walker (FLRW) metric, which forms the basis of the standard cosmological model. Observationally, the remarkable uniformity of the cosmic microwave background (CMB) has long provided strong support for isotropy since its discovery \citep{penzias65, smoot96}. Additional confirmation has come from a wide range of observations, including gamma-ray bursts \citep{meegan92}, supernovae \citep{gupta10}, radio sources \citep{blake02}, X-ray background \citep{wu99}, galaxy surveys \citep{gibelyou12, sarkar19, camila24}, and galaxy clusters \citep{bengaly17}.

Despite its success, the Cosmological Principle has come under increasing scrutiny in recent years \citep{aluri22}. Observational studies using large-scale structure tracers have reported potential deviations from statistical isotropy. These include the detection of bulk flows in Type Ia supernovae \citep{colin19}, anomalously large dipoles in the angular distribution of quasars \citep{secrest21, secrest22, kothari22}, hemispherical asymmetries in galaxy surveys \citep{wiegand14, appleby22}, and the discovery of enormous cosmic structures such as large quasar groups \citep{clowes13}, their correlated orientations \citep{friday22}, and giant cosmic voids \citep{keenan13} all of which challenge the assumption of isotropy at the largest scales. While many of these anomalies remain under active investigation and debate, they collectively underscore the need for more robust, high-sensitivity tests of isotropy across a broad range of cosmic epochs. Continued testing of the Cosmological Principle is essential, as any confirmed violation could point to new physics or expose fundamental limitations of our current cosmological framework.

Quasars are extremely luminous and distant active galactic nuclei that provide a uniquely powerful probe for such studies. Due to their brightness, they are observable across vast comoving volumes, enabling anisotropy tests at Gpc scales and high redshifts. Their spatial distribution offers a window into the evolving structure of the Universe well beyond the reach of typical galaxy surveys. In particular, quasars trace the underlying matter density field and large-scale structure in the redshift range where cosmic evolution is most dynamic, the era of peak structure formation and the onset of accelerated expansion.

However, measuring isotropy with quasars poses challenges. Their selection functions are complex and redshift-dependent, and their sky distribution is sensitive to both observational systematics and cosmic variance. To distinguish genuine anisotropies from artifacts of survey geometry or completeness, any statistical test must be both robust and sensitive to a wide range of structural scales.

In this work, we approach the problem from a new direction. We apply an information-theoretic framework based on Renyi entropy, a generalization of Shannon entropy, to assess the angular isotropy of the quasar distribution. This approach is sensitive not just to average clustering, but to the higher-order structure of the angular probability distribution making it particularly suited to identifying subtle or scale-dependent anisotropies. Using the Gaia-unWISE quasar catalog, which contains over one million uniformly selected quasars, we divide the sample into three redshift bins spanning $0.36 \leq z \leq 2.8$ and evaluate the Renyi entropy as a function of radial distance across these bins. By comparing entropy-based diagnostics with randomized mock catalogues, we quantify the emergence and disappearance of anisotropic patterns in the sky.

Our goal is to test not only whether the Universe is isotropic today, but whether isotropy emerged gradually or episodically as cosmic structure evolved. By tracing the angular structure of quasars across redshift and scale, we aim to identify potential transitional epochs, the periods in cosmic history when anisotropic features were prominent before being suppressed by the accelerating expansion of the Universe. This approach, we argue, opens a new window into the interplay between cosmic structure, entropy, and isotropy.

\section*{A new entropic lens on structure}
To uncover subtle departures from isotropy in the distribution of quasars, we need tools that go beyond traditional statistical approaches. Standard cosmological analyses often rely on correlation functions or power spectra, which quantify how density fluctuations relate across pairs of points. While powerful, these methods are limited to second-order statistics and are less sensitive to complex, non-Gaussian features that can arise during the hierarchical growth of cosmic structures. If the angular sky contains filamentary, clustered, or void-like patterns that deviate from randomness in more intricate ways, we need measures capable of capturing such higher-order structures.

This is where Renyi entropy \citep{renyi61} offers a compelling advantage. Originally introduced as a generalization of Shannon entropy, Renyi entropy provides a flexible measure of how information or structure is distributed within a system. It does so by introducing a free parameter, denoted by $q$, which controls the entropy’s sensitivity to fluctuations in the probability distribution. When $q = 1$, Renyi entropy reduces to the familiar Shannon entropy. For $q > 1$, it becomes increasingly sensitive to overdensities. By evaluating entropy across multiple values of $q$, we can probe the structure of the quasar sky at different statistical depths.

In practice, we divide the sky into equal-area pixels using the HEALPix scheme \citep{gorski05} and count the number of quasars in each pixel by including all sources within a given cumulative comoving distance. From these counts, we construct a probability distribution over the sky and compute the Renyi entropy at several values of $q$. If the distribution is isotropic and featureless, the entropy values for different $q$ will be similar and converge. In contrast, if there is structure such as clustering in one direction or a void in another, the entropies will diverge, depending on how strongly they respond to those features.

To quantify this divergence, we introduce the normalized entropy dispersion, which measures the relative spread in Renyi entropy values across orders $q$. A low dispersion indicates statistical isotropy, while a high dispersion suggests the presence of anisotropic features. This approach is sensitive not only to where the quasars are, but how their distribution departs from uniformity in ways that affect clustering, coherence, and asymmetry.

Crucially, this entropy-based method does not rely on assumptions about Gaussianity or linearity in the data. It operates directly on the spatial probability distribution, making it robust to noise and less sensitive to the particular choice of angular multipoles, which can be distorted by sky masks or irregular survey geometries. Furthermore, by comparing the measured entropy dispersion with mock realizations constructed from randomized angular positions, we can estimate the statistical significance of any detected anisotropy.

In this way, Renyi entropy acts as a lens through which cosmic structure is viewed not just as a set of positions, but as an evolving information field. It allows us to capture the richness of the quasar sky in a way that complements conventional analyses opening the door to detecting subtle, multiscale deviations from isotropy that may reflect deep features of cosmic evolution.

\section*{Detecting anisotropy with cumulative quasar distributions}
To investigate the emergence and evolution of cosmic anisotropy, we apply our entropy-based method to the angular distribution of quasars integrated cumulatively within increasing comoving distances. At each distance, we consider the complete angular distribution of all quasars enclosed within that radius, treating it as a projection of the evolving three-dimensional structure of the Universe. This approach allows us to assess how isotropy changes with enclosed volume and cosmic time, revealing the development or suppression of anisotropic features across the expanding cosmic horizon.

We use the Gaia-unWISE quasar catalogue, which provides a wide-area, photometrically selected sample of over one million quasars with high angular completeness (\autoref{fig:entire}) and well-characterized redshifts. To ensure consistency, we limit our analysis to a carefully masked region of the sky where observational quality is reliable. The sky is pixelated using the HEALPix (Hierarchical Equal Area isoLatitude Pixelation) scheme, which enables uniform angular binning across the sphere. Each pixel corresponds to a small equal-area patch of the sky, and we count the number of quasars falling within each pixel to construct angular density maps.

\begin{figure*}[htbp!]
\centering\includegraphics[width=10cm]{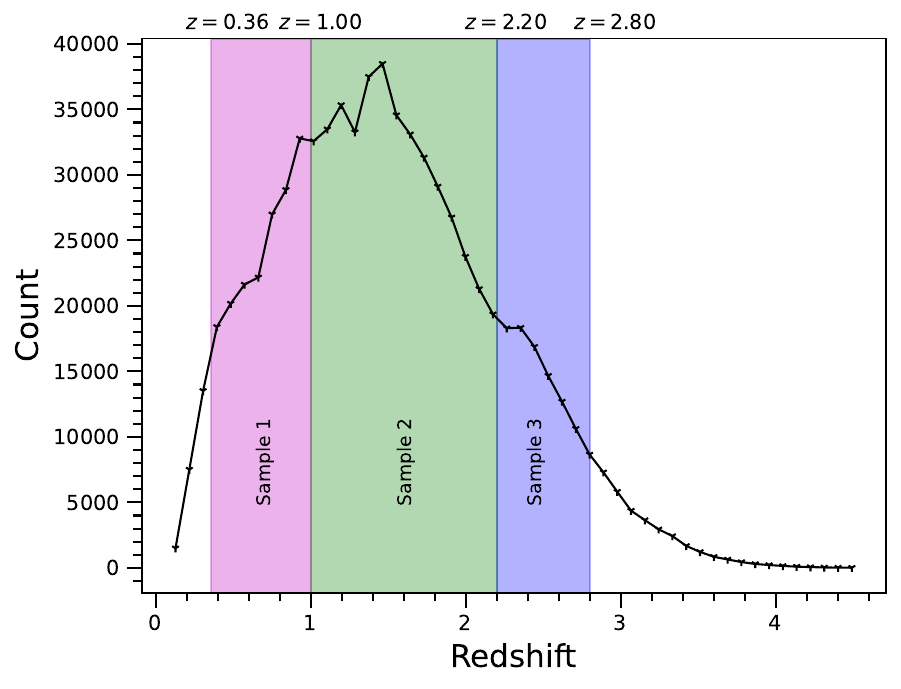}
\centering\includegraphics[width=10cm]{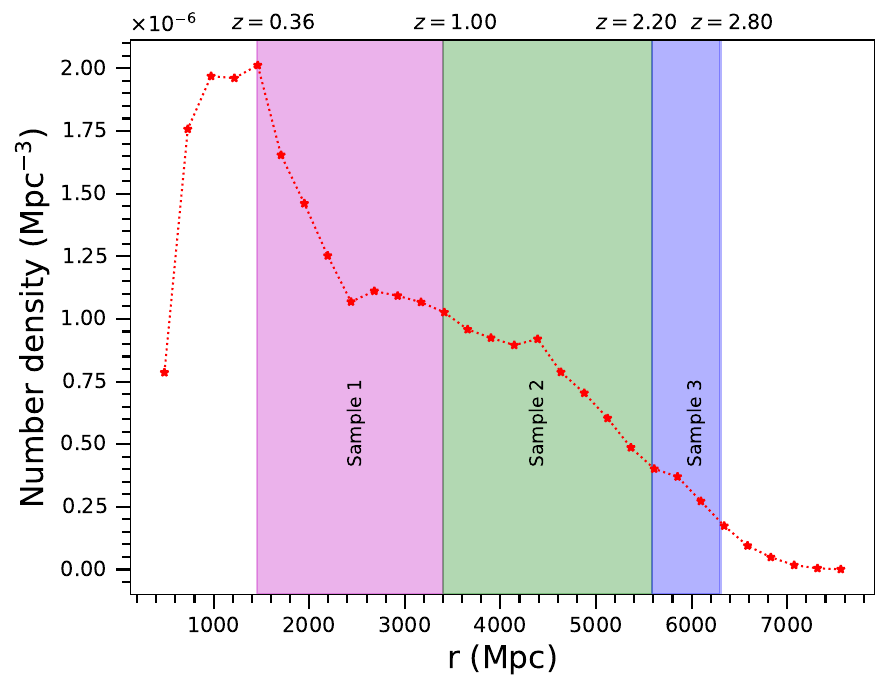} 
\caption{The top panel shows the distribution of quasar counts as a function of redshift, while the bottom panel displays the variation of quasar number density with comoving radial distance. The three redshift samples used in our analysis are highlighted in different colours, Sample 1 ($0.36 \leq z < 1.00$) in magenta, Sample 2 ($1 \leq z < 2.2$) in green, and Sample 3 ($2.2 \leq z \leq 2.8$) in blue.}
\label{fig:zbins}
\end{figure*}

The quasar sample is divided into three redshift intervals: Sample~1 ($0.36 \leq z < 1$), Sample~2 ($1 \leq z < 2.2$), and Sample~3 ($2.2 \leq z \leq 2.8$) (\autoref{fig:zbins}, \autoref{tab:table1}). These bins are chosen to balance number statistics, completeness, and cosmological relevance covering the post-reionization structure formation era, the peak of large-scale clustering, and the transition into dark energy domination. Within each redshift bin, we track how the angular distribution evolves by incrementally including all quasars within increasing comoving distances from the observer. 

For each cumulative comoving radius, we compute the Renyi entropy across multiple orders $q$ using the pixelized angular distribution. We then calculate the normalized entropy dispersion, which quantifies the spread among entropy values across $q$ and serves as a sensitive indicator of angular anisotropy. A high dispersion suggests scale-dependent structural patterns, while low values indicate statistical isotropy.

To evaluate the significance of the observed anisotropies, we generate $100$ isotropic mock realizations separately for each redshift bin, accounting for the slight differences in the masked quasar sky due to completeness-based masking applied independently to each sample. In these mocks, the radial distances of the quasars are preserved, but their angular coordinates are randomized uniformly within the unmasked sky. This process maintains the redshift-dependent number density and survey geometry while eliminating any real or artificial anisotropy. Entropy dispersions are calculated for each mock using the same cumulative procedure as for the data.

We define a significance ratio $\Psi(r)$ as the difference between the observed and mock mean entropy dispersions, normalized by the mock standard deviation. This statistic quantifies the deviation from isotropy at each comoving distance in units of statistical uncertainty.

The analysis is conducted under two masking strategies: a standard mask that excludes regions near the Galactic plane and areas of poor coverage (\autoref{fig:mask1}), and a more stringent mask that additionally removes regions likely to be affected by photometric systematics or stellar contamination (\autoref{fig:mask2}). Consistent results across these masks provide confidence in the robustness of the detected signals.

It is important to note that the quasar samples used in our analysis span wide redshift ranges and therefore do not represent the matter distribution on a single cosmic time slice. Instead, they trace an evolving density field shaped by gravitational clustering over extended epochs.  However, since we integrate the quasar distribution within each HEALPix pixel up to a fixed comoving radial distance, this approach enables meaningful comparisons across pixels for testing isotropy, even though the underlying structure evolves with time.

Through this cumulative approach, each angular distribution acts as a projection of the Universe’s structure up to a given comoving radius. By tracing how entropy-based measures evolve with scale, we can identify epochs where anisotropy emerges, peaks, or diminishes providing insight into the timing and dynamics of large-scale structure formation and the influence of cosmic acceleration.

\section*{A transitional epoch emerges}
Our analysis reveals a striking and statistically significant signature of anisotropy in the quasar distribution but only during a specific epoch of cosmic time. Of the three redshift bins analysed, the intermediate range ($1 \leq z < 2.2$) stands out with a clear and coherent departure from statistical isotropy (\autoref{fig:dispersion1}). In contrast, the lower ($0.36 \leq z < 1$) and higher ($2.2 \leq z \leq 2.8$) redshift intervals exhibit entropy statistics consistent with isotropic expectations (\autoref{fig:dispersion1}). This sharp contrast suggests that anisotropy is not a pervasive feature of the quasar sky, but rather one that emerges during a transitional epoch in the evolution of cosmic structure.

In the intermediate redshift bin, the significance ratio $\Psi(r)$ which quantifies deviations in entropy dispersion relative to isotropic mocks rises monotonically with comoving distance, reaching values as high as $12.5\sigma$ at scales near $5600$ Mpc (\autoref{fig:dispersion1}). The trend persists with reduced but high significance ($\sim 5\sigma$) in \autoref{fig:dispersion2} even under a conservative mask (\autoref{fig:mask2}) that excludes regions potentially affected by photometric systematics or Galactic contamination. Although some fluctuation in entropy dispersion is expected from shot noise and survey geometry, the magnitude and coherence of the signal in this redshift range exceed the variance seen in any of the randomized mock realizations.

A particularly striking feature in our analysis is the steadily increasing deviation from isotropy observed in Sample~2 between comoving radial distances of $4500$ Mpc and $5600$ Mpc. This trend, absent in the other redshift bins, signals a transitional epoch in cosmic evolution, an era during which large-scale structures such as superclusters, filaments, and voids were rapidly forming through gravitational instability. These coherent patterns introduce directional inhomogeneities in the quasar distribution, which manifest as measurable anisotropies in the angular sky. The entropy signature we detect in this radial range reflects the peak of matter-driven structure growth, a moment when cosmic complexity briefly surged before being gradually suppressed by the onset of accelerated expansion. At lower redshifts, dark energy slows the formation of new structures, leading to a statistical reemergence of isotropy. Thus, the observed anisotropy in Sample~2 (\autoref{fig:dispersion1} and \autoref{fig:dispersion2}) provides a physically compelling and observationally robust imprint of a transitional phase in the evolution of cosmic structure.

This behaviour stands in stark contrast to the low- and high-redshift bins. In the low-redshift range, where cosmic acceleration dominates and the number density of quasars is lower, entropy dispersions are small and $\Psi(r)$ remains close to unity which is fully consistent with statistical isotropy. Similarly, the high-redshift bin shows no significant anomalies, likely reflecting the early, nearly homogeneous conditions of the Universe when structure had not yet grown to appreciable scales.

Moreover, this transition aligns with a recent theoretical proposal that link the dynamical behaviour of the Universe to the evolution of its configuration entropy. If the gravitational clustering of matter leads to a dissipation of entropy, the Universe may respond by accelerating its expansion to stabilize or restore the balance of information \citep{pandey17}. The entropy-driven anisotropy we observe could therefore serve not just as a tracer of structure formation, but as a potential driver of cosmological dynamics itself.

In summary, our findings point to a distinct transitional epoch in the cosmic timeline an interval of enhanced anisotropy bounded by symmetry at both earlier and later times. This discovery supports a dynamic view of cosmic isotropy and opens the door to new connections between structure, entropy, and the expansion history of the Universe.

\section*{Conclusions}
The detection of a significant anisotropy confined to a specific redshift interval raises a profound question: what does this transitional behaviour reveal about the evolution of the Universe? The entropy signal we detect is not random nor persistent across time, but sharply bounded in a cosmologically meaningful epoch emerging strongly at $1 \leq z < 2.2$ and fading on either side. This pattern hints at an underlying dynamical mechanism tied to the growth and suppression of cosmic structure.

The intermediate-redshift interval corresponds to the peak era of large-scale structure formation. During this epoch, gravitational instability had matured the cosmic web into filaments, voids, and superclusters, the features capable of imprinting coherent angular anisotropies in the quasar sky. At higher redshifts, the Universe was closer to statistical homogeneity, while at lower redshifts, the accelerating expansion driven by dark energy began suppressing further growth of large-scale structures.

A key interpretive question concerns the observed restoration of isotropy at $z < 1$. Since we evaluate Renyi entropy in comoving coordinates, one might expect that anisotropies formed at earlier times should persist, even after structure growth freezes. However, our results show a clear decline in entropy dispersion at low redshifts, a trend that is not due to geometric stretching but results from a combination of observational and astrophysical factors. The total number of quasars per shell decreases due to the smaller comoving volume in Sample~1, increasing shot noise and reducing contrast. Additionally, the reduced shell volume limits the number of structures sampled. Most importantly, the onset of accelerated expansion halts the imprinting of new anisotropies. While Sample~1 lies entirely before the epoch where dark energy dominates the cosmic energy budget ($z \sim 0.3$), the suppression of structure growth begins earlier as early as $z \sim 1$ when dark energy starts to affect the expansion rate \citep{bernal16, garcia21, nguyen23, abbott23}. The lack of significant anisotropy in Sample~1 is therefore consistent with a Universe where the growth of complexity has already slowed, and new anisotropies cease to emerge.

This interpretation aligns with thermodynamic perspectives on cosmic evolution \citep{radicella12, pandey17, pandey19a, pandey19b}. As matter clusters, the configuration entropy of the Universe decreases. If unbalanced, this entropy loss would conflict with the second law of thermodynamics. One proposed resolution is that the Universe dynamically responds via accelerated expansion, halting further structure formation to preserve entropy balance \citep{pandey17}. Our results showing entropy dispersion rising and then falling could reflect this thermodynamic feedback: a phase of complexity growth followed by accelerated expansion.

While our analysis includes rigorous mock-based significance estimates and robust masking to exclude low-quality regions, we acknowledge that the role of residual systematics cannot be fully ruled out. Possible sources include redshift-dependent variations in quasar selection efficiency \citep{fan99}, spatially correlated photometric uncertainties \citep{huterer13}, and incomplete removal of galactic contamination \citep{kothari22}. Nevertheless, the fact that the anisotropy signal emerges only in the intermediate redshift bin, and is absent in both the lower and higher bins, argues against a purely systematic origin. A genuine cosmological signal remains the most plausible explanation. 

Confirming this conclusion will require cross-validation with independent quasar catalogues and alternative tracers of large-scale structure. In future work, we plan to apply Renyi entropy-based techniques to upcoming datasets from the Large Synoptic Survey Telescope (LSST) \citep{lsst19}, Euclid \citep{euclid25}, and the Dark Energy Spectroscopic Instrument (DESI) \citep{desi24}, which will offer unprecedented depth and statistical power. These surveys will enable new tests of isotropy across redshift, angular scale, and tracer type, helping to refine our understanding of cosmic structure and its evolution.

In conclusion, we identify a transitional epoch of cosmic anisotropy in the angular distribution of quasars, revealed through Renyi entropy applied to the cumulative distribution of quasars within increasing comoving distances. This anisotropy remains statistically robust under variations in HEALPix resolution, masking strategy, and the number of randomized mock realizations used in the analysis (\autoref{fig:64mask1}, \autoref{fig:64mask2}, \autoref{fig:converge}).

Rather than contradicting the Cosmological Principle, our findings illustrate how large-scale isotropy can emerge dynamically from the projected imprint of an evolving cosmic web, shaped by gravitational clustering and modulated by the onset of cosmic acceleration. Renyi entropy thus opens a new observational window into the structure of the Universe, capturing how complexity grows, saturates, and eventually smooths out across the expanding cosmic horizon.

\bibliographystyle{aasjournal}
\bibliography{refs.bib}

\section*{Methods}
\begin{figure*}[htbp!]
\centering\includegraphics[width=12cm]{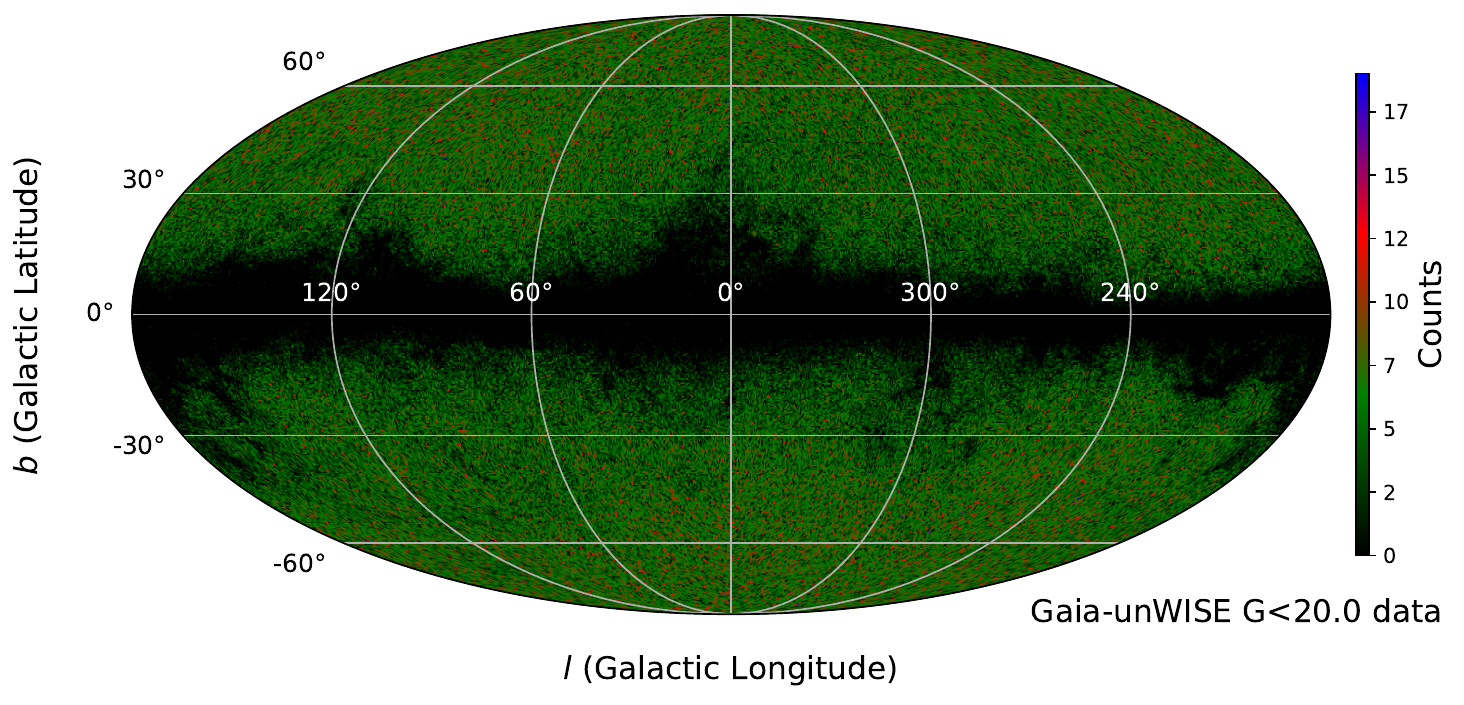}
\caption{Sky map of quasar counts from the Gaia-unWISE catalog ($G < 20.0$), shown in Galactic coordinates using HEALPix pixelization with resolution parameter $N_{\mathrm{side}} = 128$.}
\label{fig:entire}
\end{figure*}

\subsection*{Data and Catalogues}
We use the publicly available Gaia–unWISE quasar catalogue\citep{storey24}, a nearly all-sky quasar compilation that combines optical data from Gaia DR3 \citep{gaia23a, gaia23b} with mid-infrared photometry from WISE \citep{wright10}, reprocessed through the unWISE pipeline \citep{lang14, meisner19}. To ensure photometric uniformity and minimize dust extinction, the catalogue excludes the Galactic disk, yielding a clean sky coverage of 30,277.52 deg$^2$ approximately $73\%$ of the entire sky. Among existing quasar datasets, Gaia-unWISE probes the largest comoving volume with high angular completeness. The catalogue begins with 6,649,162 quasar candidates from Gaia DR3, cross-matched with unWISE data in the W1 and W2 infrared bands. A series of proper motion filters and color-magnitude cuts are applied to reduce stellar contamination and enhance sample purity. Two main subsamples are provided based on Gaia $G$-band magnitude limits: one with $G<20.5$ (1,295,502 sources) and a cleaner, more conservative subsample with $G<20.0$ (755,850 sources). We adopt the latter in our analysis and divide the data into three redshift bins for investigating anisotropy across cosmic epochs.

For this analysis, we limit the sample to the redshift range $0.36 \leq z \leq 2.8$. The quasar sample is divided into three redshift bins: Sample~1 ($0.36 \leq z < 1$), Sample~2 ($1 \leq z < 2.2$), and Sample~3 ($2.2 \leq z \leq 2.8$), containing 51,320, 124,924, and 30,261 quasars, respectively. Each redshift bin is subdivided into 30 cumulative comoving intervals to facilitate a uniform comparison of entropy evolution across cosmic time. The thickness of each radial shell is adapted to the number density and total count of quasars in the respective bin: 64.663\,Mpc for Sample~1, 73.115\,Mpc for Sample~2, and 23.74\,Mpc for Sample~3. These values correspond to number densities of $1.83 \times 10^{-6}$, $7.33 \times 10^{-7}$, and $3.18 \times 10^{-7}$ quasars per Mpc$^3$, respectively (see \autoref{tab:table1}). The smaller shell thickness in Sample~3 compensates for its lower number density and smaller total sample size, thereby maintaining sufficient angular resolution and statistical power for the entropy calculations at each scale. The comoving distance ranges for the three samples are listed in \autoref{tab:table1}. We use a flat $\Lambda$CDM cosmology with parameters consistent with Planck 2018 results \citep{planck18}.

\begin{table*}[htbp]
\centering
\caption{Summary of quasar samples.}
\label{tab:table1} 
\resizebox{\textwidth}{!}{
\begin{tabular}{|c|c|c|c|c|c|c|}
\hline
Quasar sample & Redshift range & \makecell{Radial extension \\ (Mpc)} & \makecell{Quasar\\count} & \makecell{Mean density\\($\text{Mpc}^{-3}$)} & \makecell{Quasar count\\after applying \\ Mask~1} & \makecell{Quasar count\\after applying \\ Mask~2} \\
\hline
Sample 1 & $0.36 \leq z < 1$ & $1455.74 \leq r < 3395.63$ & 178785 & $1.83\times10^{-6}$ & 51320 & 40894 \\
\hline
Sample 2 & $1 \leq z < 2.2$ & $3395.63 \leq r < 5589.08$ & 415724 & $7.33\times10^{-7}$ & 124924 & 99737 \\
\hline
Sample 3 & $2.2 \leq z \leq 2.8$ & $5589.08 \leq r \leq 6301.48$ & 100637 & $3.18\times10^{-7}$ & 30261 & 24590 \\
\hline
\end{tabular}
}
\end{table*}

\subsection*{Sky masking and survey geometry}
\begin{figure*}[htbp!]
\centering\includegraphics[width=12cm]{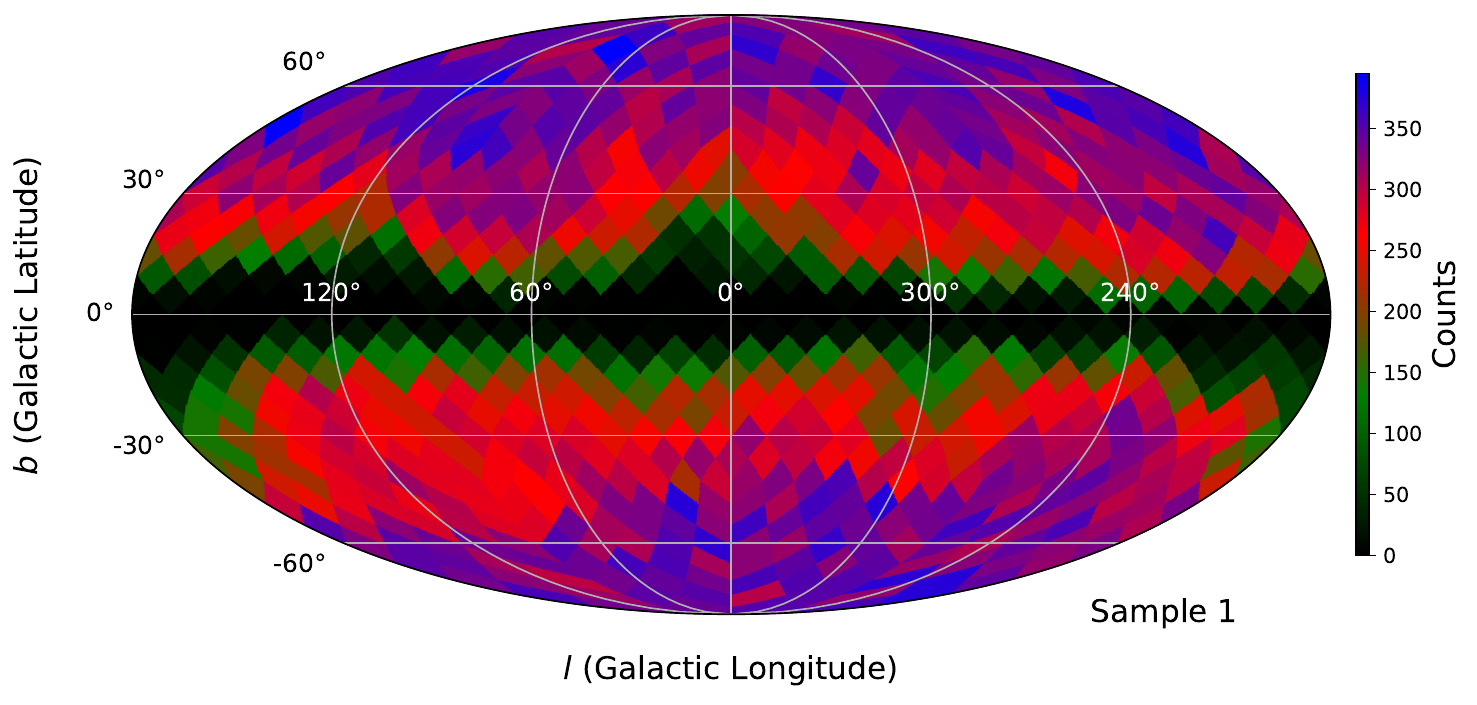}
\centering\includegraphics[width=12cm]{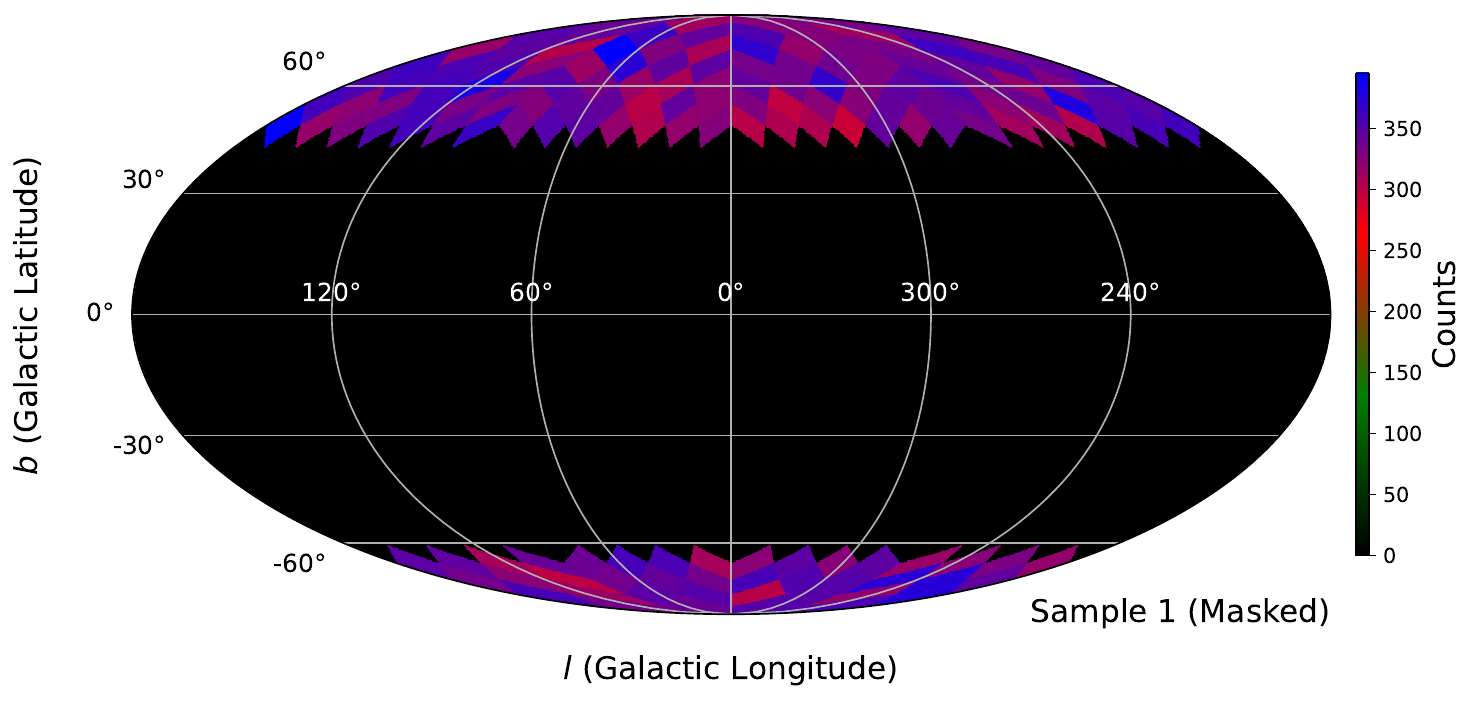}
\centering\includegraphics[width=12cm]{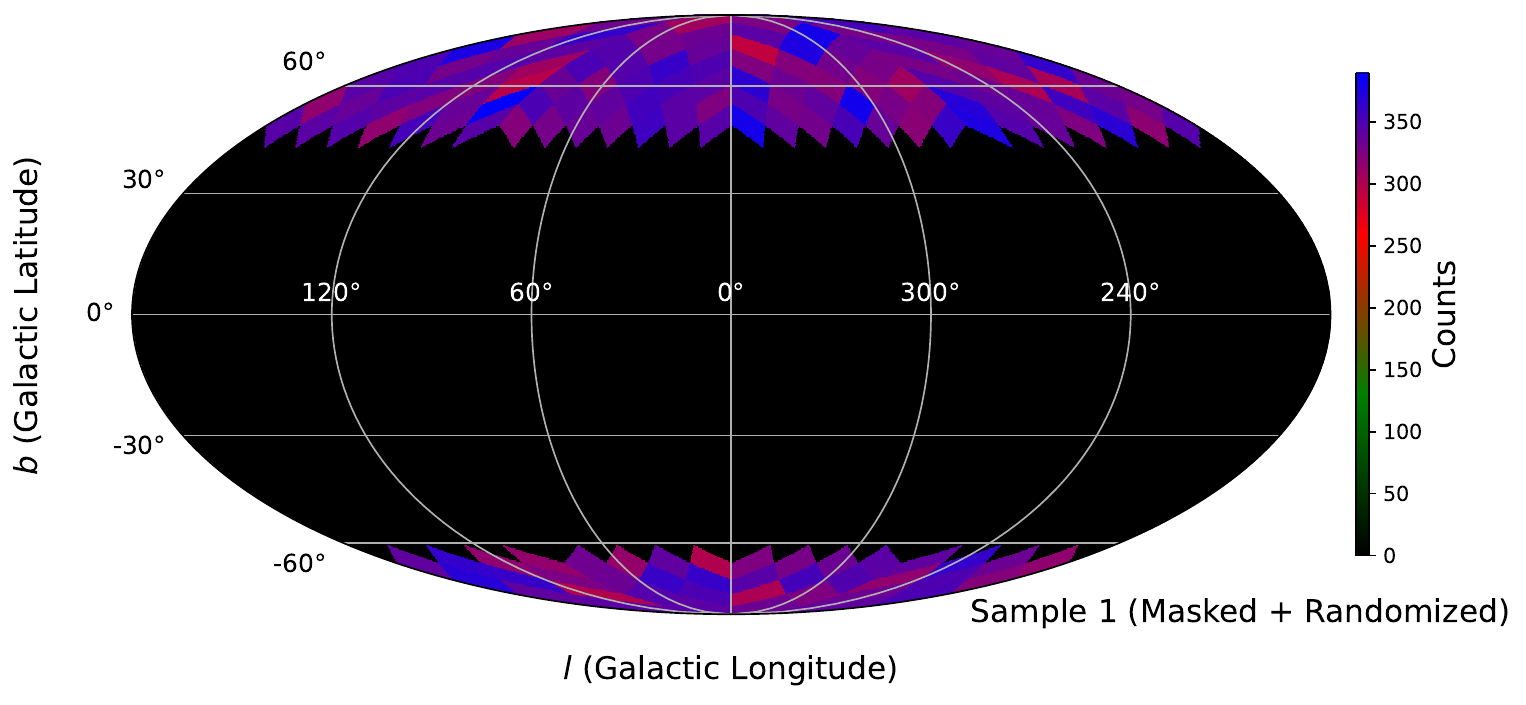}
\caption{Quasar count maps for Sample 1, constructed using HEALPix with resolution parameter $N_{\mathrm{side}} = 8$. The top panel shows the raw quasar distribution.  The middle panel displays the same sample after applying Mask 1 to exclude regions with potential systematics. The bottom panel shows a randomized version of Sample 1 after masking, where angular positions of the quasars have been shuffled while preserving their radial distances.}
\label{fig:mask1}    
\end{figure*}

\begin{figure*}[htbp!]
\centering\includegraphics[width=12cm]{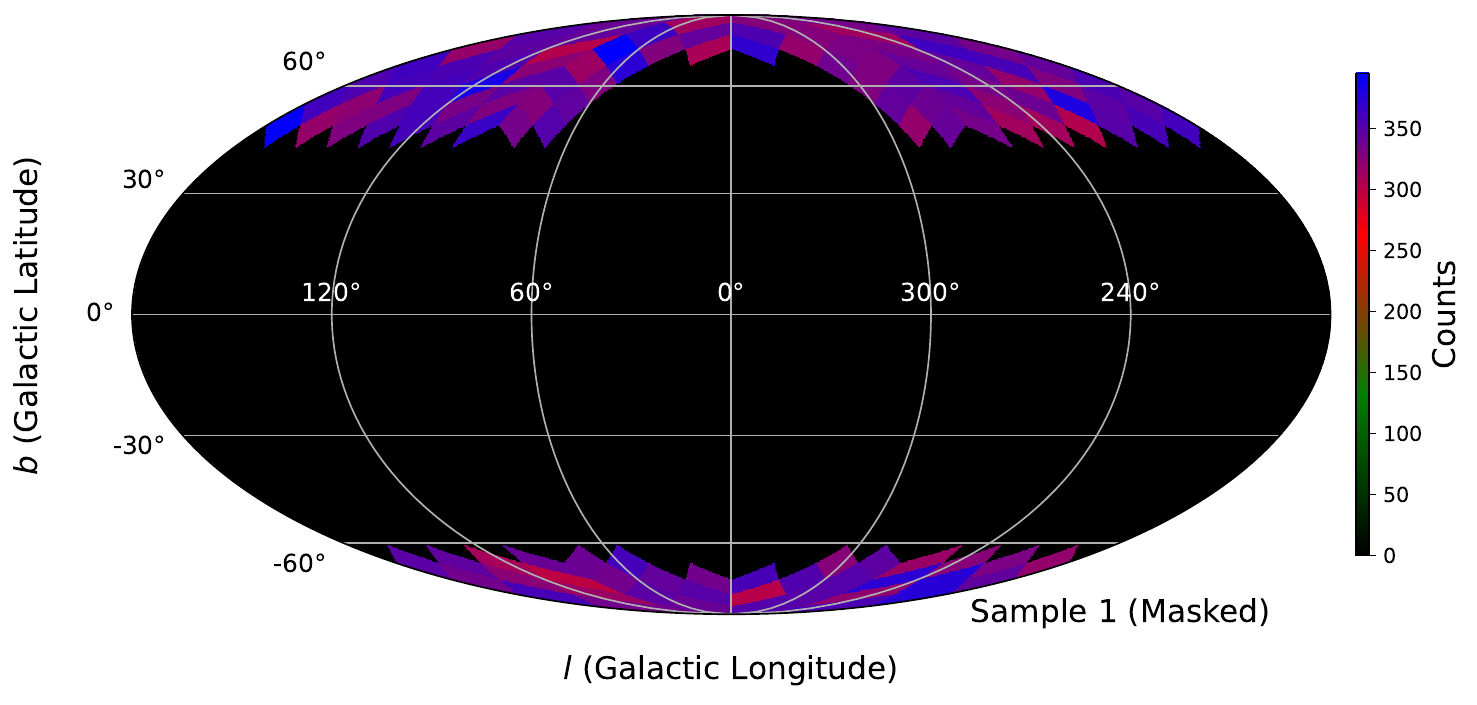}
\centering\includegraphics[width=12cm]{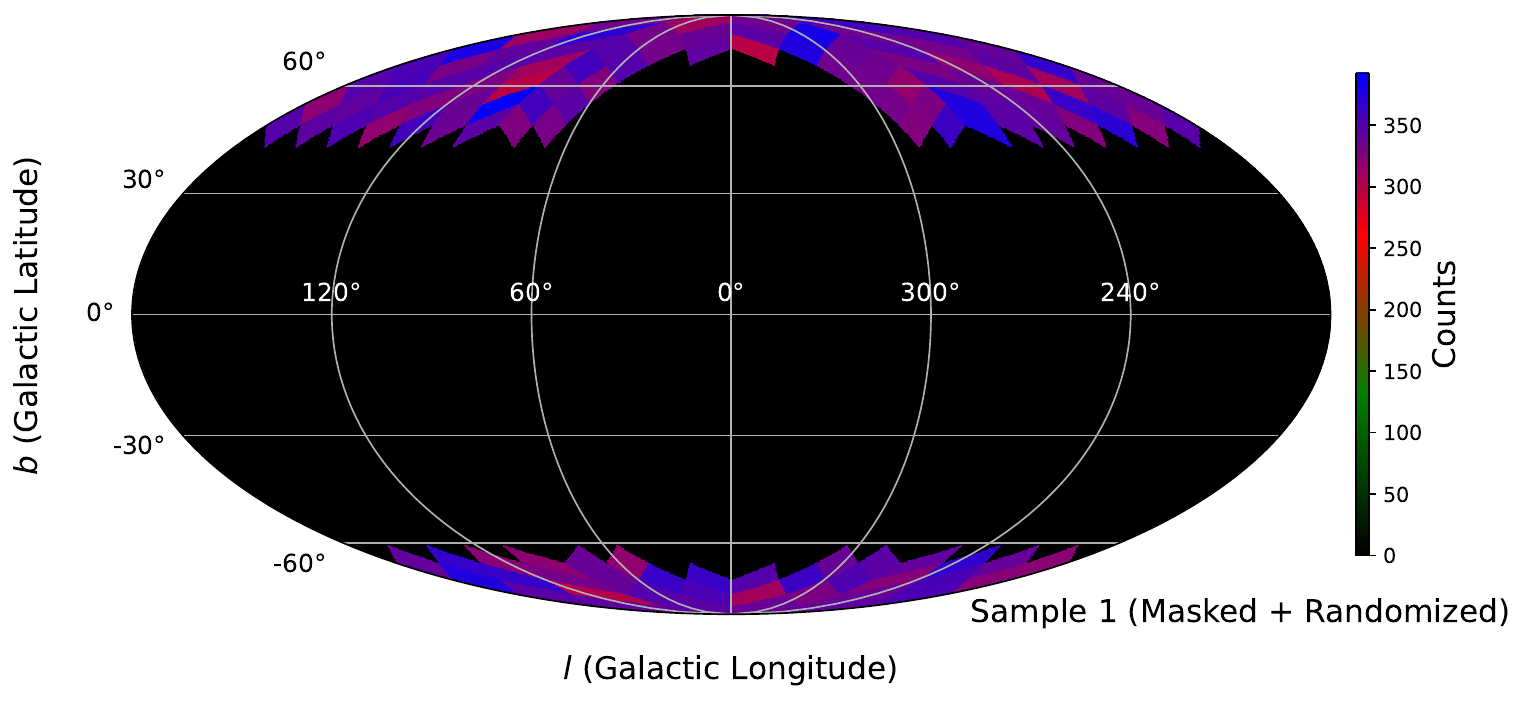}
\caption{The top panel displays the quasar counts for Sample 1 after applying a stringent mask (Mask 2) that excludes regions with potential systematics.  The bottom panel shows a randomized version of Sample 1 after masking.}
\label{fig:mask2}    
\end{figure*}

\begin{figure*}[htbp!]
\centering\includegraphics[width=14cm]{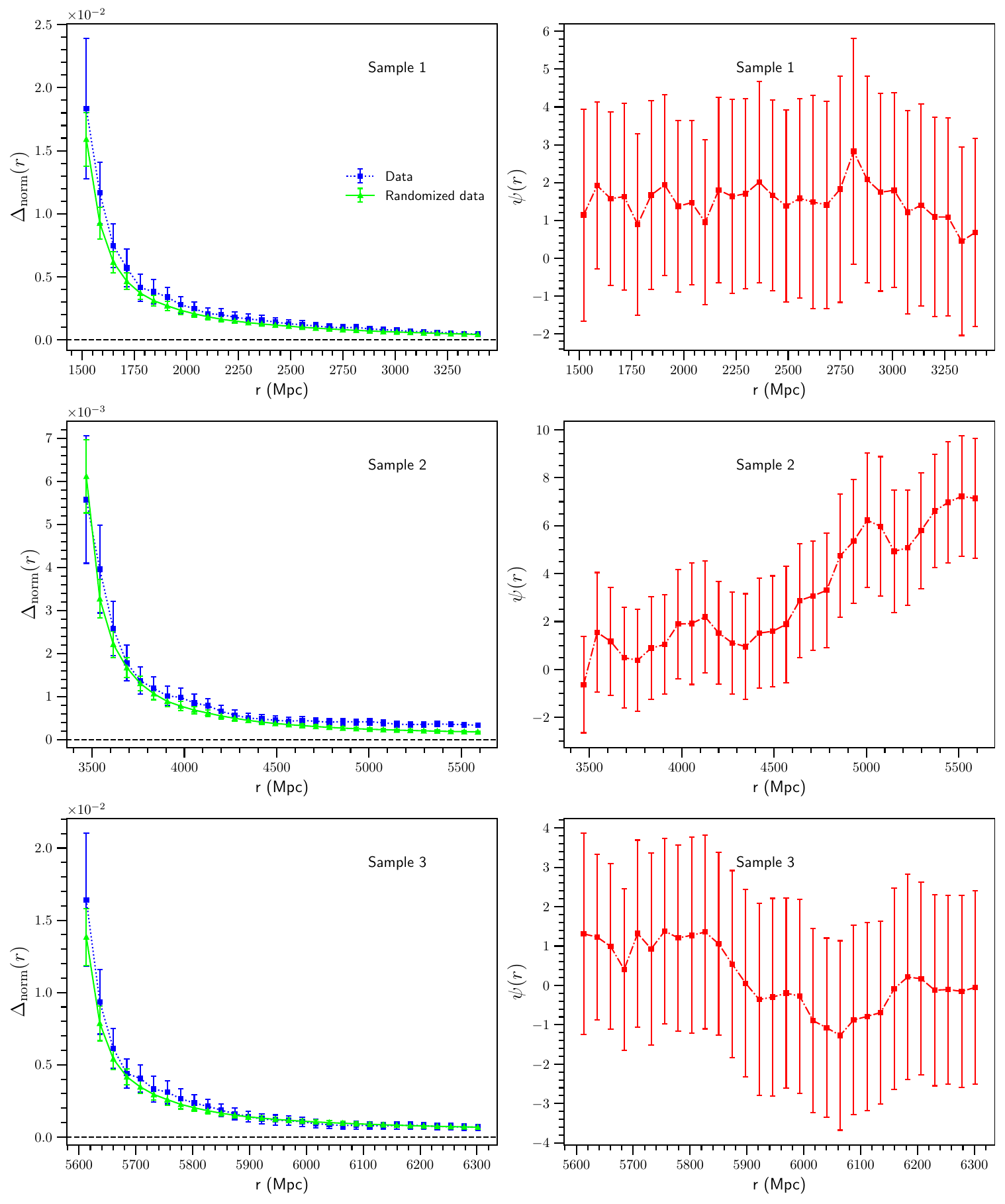}
\caption{Same as \autoref{fig:mask1} but when using a stringent mask that excludes regions with potential systematics.}
\label{fig:dispersion2}
\end{figure*}

To minimize the impact of Galactic contamination and survey inhomogeneities, we employ two distinct sky masks in our analysis: a standard mask to exclude the Galactic plane and poorly observed regions (\autoref{fig:mask1}), and a conservative mask that removes additional areas affected by stellar contamination or photometric systematics (\autoref{fig:mask2}). These masks are designed to ensure uniform data quality, suppress spurious angular features, and enhance the reliability of our anisotropy measurements.

\subsubsection*{Mask 1: Sky pixel completeness-based mask}
Our first masking strategy is based on angular completeness within HEALPix-defined regions. We begin by pixelating the full sky using the HEALPix scheme \citep{gorski05}, which divides the celestial sphere into equal-area pixels. For resolution parameter $N_{\mathrm{side}} = 8$, the sky is divided into $12 \times N_{\mathrm{side}}^2 = 768$ coarse ``super-pixels.''

To remove areas with poor quasar sampling, we use a hierarchical masking approach. Each super-pixel at $N_{\mathrm{side}} = 8$ is subdivided into 64 sub-pixels at $N_{\mathrm{side}} = 64$. We then calculate the fraction of these sub-pixels that contain at least one quasar. If less than 90\% of a super-pixel’s sub-pixels are populated (i.e., if the filled fraction $f < 0.9$), the entire super-pixel is excluded from the analysis.

This pixel-based mask is applied only after excluding regions with high Galactic extinction. Specifically, we retain only quasars with Galactic latitude $b \geq 40^\circ$ or $b \leq -60^\circ$. These thresholds are chosen to avoid the dusty Galactic plane and bulge, where photometric systematics and extinction significantly degrade data quality.

\autoref{fig:entire} shows the full-sky quasar count map for the $G<20.0$ sample at $N_{\mathrm{side}} = 128$, illustrating the large-scale angular coverage and uniformity of the dataset. \autoref{fig:mask1} (top panel) displays the lower-resolution count map for Sample~1 at $N_{\mathrm{side}} = 8$, highlighting the underdensity near the celestial equator. After applying our hierarchical completeness mask, the resulting sky coverage is shown in the middle panel of \autoref{fig:mask1}, reflecting the regions retained for isotropy analysis.

\subsubsection*{Mask 2: Circular cut mask}
To further guard against residual systematics particularly those related to photometric artifacts and stellar contamination we introduce a second, more conservative masking strategy based on a geometric sky cut.

This circular mask is centered at the Galactic coordinates $(l = 0^\circ, b = 0^\circ)$, corresponding to the Galactic center, where both extinction and stellar density are highest (\autoref{fig:mask2}). The goal is to exclude a circular region that subtends 4 steradians of solid angle, roughly one-third of the sky \citep{mittal24}.

To determine the angular radius of the exclusion zone used in our circular mask, we calculate the solid angle $\Omega$ subtended by a spherical cap on the sky. The relation between the solid angle and the angular radius $\theta_0$ of such a cap is given by $\Omega = \int_{0}^{\theta_0} \int_{0}^{2\pi} \sin \theta \,d\theta \,d\phi = 2\pi (1 - \cos \theta_0)$. Setting $\Omega = 4 \, \text{sr}$, we solve for \( \theta_0 \) as $\theta_0 = \cos^{-1}\left(1 - \frac{2}{\pi}\right) \approx 68.6^\circ$. This defines a circular region centered on the Galactic center $(l = 0^\circ, b = 0^\circ$) with angular radius $\theta_0$, which we exclude from our analysis. To implement this mask, we compute the angular distance $\delta$ of each quasar from the Galactic center using the spherical law of cosines as $\delta = \cos^{-1}(\cos l \cos b)$. All quasars with $\delta \leq 68.6^\circ$ are excluded from the analysis and only quasars with $\delta > 68.6^\circ$ are retained. This ensures that our final sample excludes a symmetric region around the Galactic center, minimizing contamination from stellar crowding and photometric systematics. This criterion is applied after the Galactic latitude mask and in conjunction with Mask~1, forming a highly conservative sky mask (\autoref{fig:mask2}) that minimizes any residual observational biases.

\subsubsection*{Masking strategy and rationale}
By using two complementary masking strategies one based on hierarchical pixel completeness and the other on geometric exclusion we are able to rigorously test the robustness of our anisotropy results. Mask~1 (\autoref{fig:mask1}) ensures statistical uniformity in angular sampling, while Mask~2 (\autoref{fig:mask2}) provides an extra layer of protection against potential contamination near the Galactic plane and center. The comparison of results under both masks allows us to assess the sensitivity of our findings to sky coverage and systematics, and reinforces the physical significance of the anisotropy signal we detect.

\subsection*{Generating isotropic mock realizations}
To establish a reference for statistical isotropy, we generate 100 randomized mock realizations based on the masked quasar data. Each mock preserves the full radial (redshift) distribution of the real sample while randomizing the angular positions of the quasars. Starting from the observed catalog, we extract the galactic coordinates longitude ($l$) and latitude ($b$) and redshift ($z$) for each quasar. These are converted into spherical polar coordinates: the comoving radial distance $r$ is derived from $z$, while the polar angle $\theta$ and azimuthal angle $\phi$ correspond to $90^\circ - b$ and $l$, respectively.

In constructing each mock, we randomize the angular coordinates $(\theta, \phi)$ uniformly across the sky, while retaining each quasar’s original comoving distance $r$. To preserve the survey geometry and sky coverage of the actual data, we restrict the randomized angular positions to fall within the same set of HEALPix pixels that are populated in the masked data, using a resolution parameter $N_{\mathrm{side}} = 8$. This ensures that the mocks respect the same sky mask and angular selection function as the data, avoiding artificial artifacts due to mismatched geometry.

The resulting mock catalogues provide statistically isotropic quasar distributions with the same radial selection function and sky coverage as the actual data. The bottom panel of \autoref{fig:mask1} displays the HEALPix count map for one such randomized mock realization of Sample~1, illustrating the uniform angular distribution over the masked sky.

\begin{figure*}[htbp!]
\centering
\includegraphics[width=14cm]{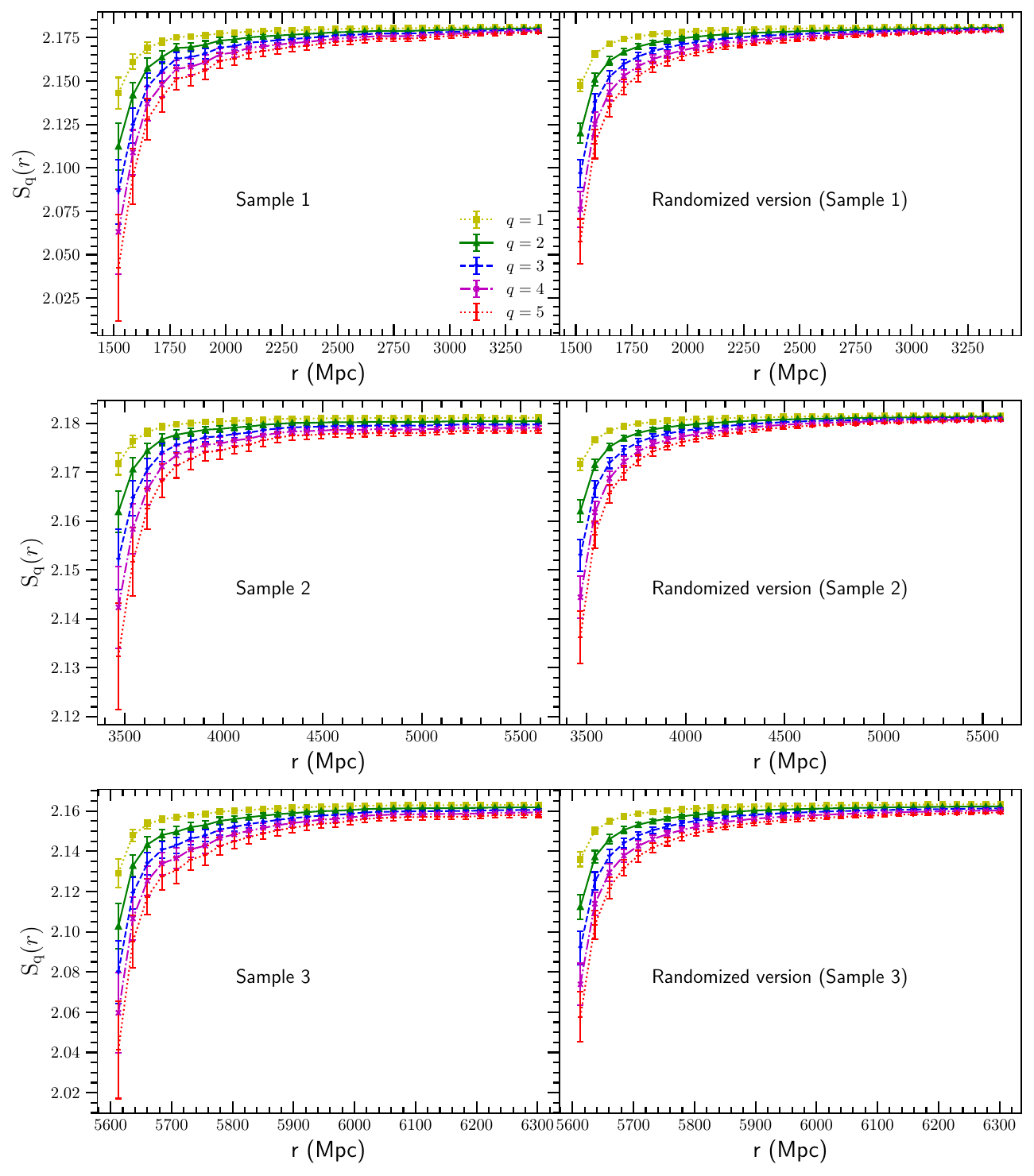}
\caption{This shows the variation of Renyi entropy $S_q(r)$ with comoving radial distance $r$ for different entropy orders $q = 1, 2, 3, 4, 5$. Left panels show results for the Gaia–unWISE quasar data and right panels show the corresponding results for isotropic mock realizations. The top, middle, and bottom panels correspond to Samples~1, 2, and 3, respectively. In each panel, the curves represent the mean Renyi entropy for different $q$ values along with the associated  1$\sigma$ uncertainty. The 1$\sigma$ error bars are estimated using 100 bootstrap resamplings for the data and 100 mock realizations for the randomized data. All entropy orders exhibit a monotonic increase with $r$. However, clear differences in the separation and convergence behaviour of the entropy curves are observed between the data and mocks, particularly for Sample~2. These differences motivate the use of normalized entropy dispersion and the significance ratio to quantify anisotropy. All results shown here are based on the application of Mask~1.}
\label{fig:combined_ent}
\end{figure*}

\begin{figure*}[htbp!]
\centering\includegraphics[width=14cm]{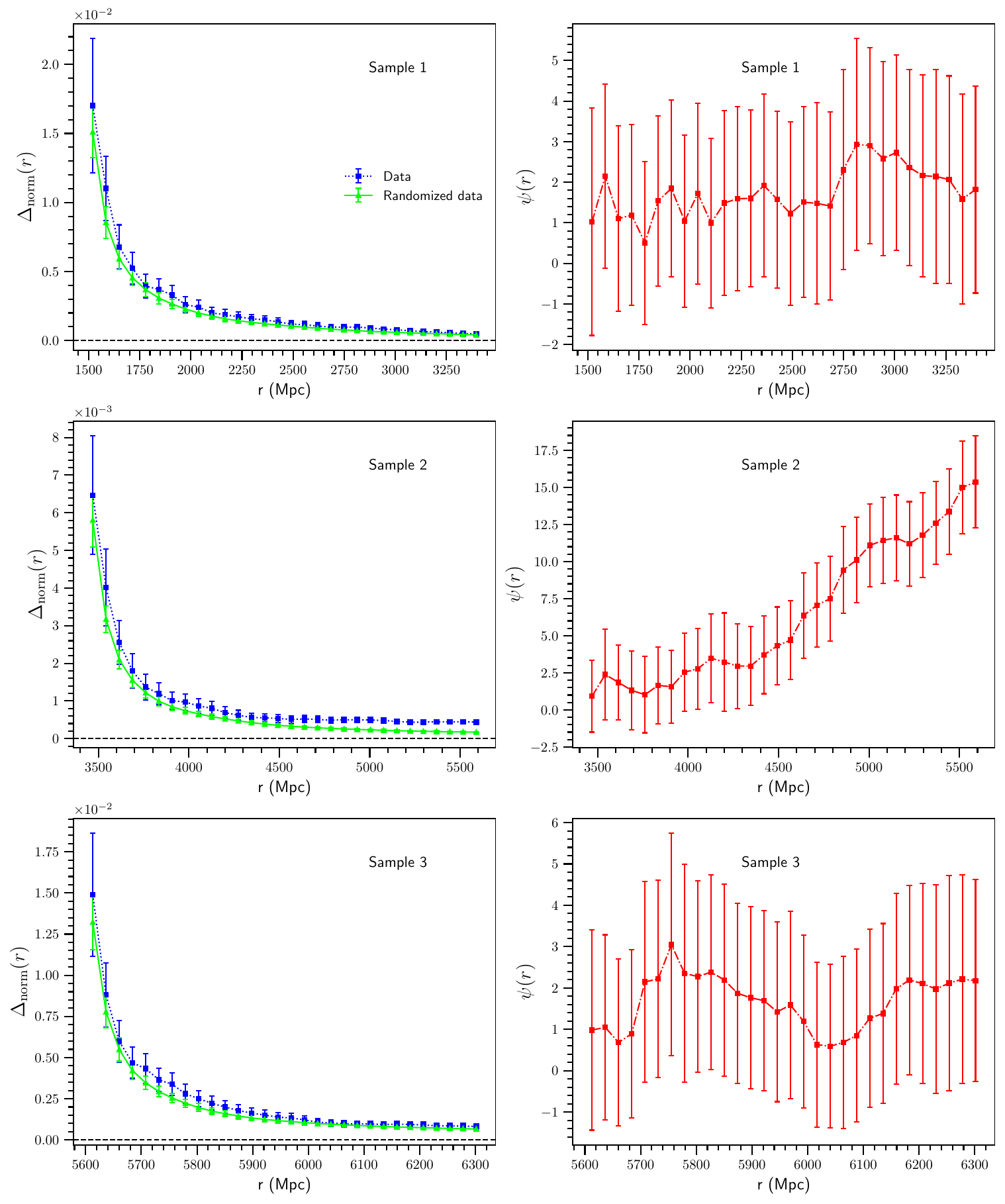}
\caption{The left panels show the mean normalized entropy dispersion $\Delta_{\mathrm{norm}}(r)$  as a function of comoving radial distance $r$, computed from the Renyi entropies of the Gaia-unWISE quasar distribution (dotted blue lines). The 1$\sigma$ error bars are derived from 100 bootstrap resamplings of the data. For comparison, the solid green lines represent the mean dispersion from 100 isotropic mock realizations, in which the radial distances of quasars are preserved but their angular positions are randomized. The corresponding 1$\sigma$ uncertainties for the mocks are estimated from the same ensemble of realizations. The right panels display the significance ratio $\Psi(r)$, indicating the deviation of the data from the isotropic expectation in units of the mock standard deviation. Uncertainties on $\Psi(r)$ are obtained via error propagation using bootstrap errors from the data and mock ensembles. Top, middle, and bottom panels correspond to Samples~1, 2, and 3, respectively.
}
\label{fig:dispersion1}
\end{figure*}

\subsection*{Renyi entropy}
Quantifying structure in the Universe requires more than measuring average density fluctuations. It calls for tools that can capture the full statistical complexity of a distribution. In information theory, this complexity is often described using entropy. The most well-known measure, Shannon entropy, introduced by Claude Shannon \citep{shannon48}, quantifies the average information content or uncertainty associated with a discrete random variable. For a variable $Y$ with $n$ discrete outcomes $\{y_1, y_2, \dots, y_n\}$, occurring with probabilities $p(y_i)$, the Shannon entropy is defined as
\begin{equation}
H(Y) = - \sum_{i=1}^{n} p(y_i) \log p(y_i).
\label{eq:shannon_entropy}
\end{equation}

While powerful, Shannon entropy treats all features of a distribution equally. To generalize this concept and make it sensitive to different aspects of structure such as the prominence of overdensities or underdensities Renyi entropy was introduced by Alfred Renyi \citep{renyi61}. The Renyi entropy of order $q$ is given by

\begin{equation}
S_q(Y) = \frac{1}{1 - q} \log \sum_{i=1}^{n} p^q(y_i),
\label{eq:renyi_entropy}
\end{equation}

where $ q \in [0, \infty] $ is a tunable parameter. For $q = 0$, the Renyi entropy simply measures the logarithm of the number of non-zero probability elements which corresponds to the maximum possible entropy. It treats all populated states as equally probable and is insensitive to the actual distribution of probabilities. As $q \rightarrow 1$, Renyi entropy reduces to Shannon entropy. When all outcomes are equally probable, Renyi entropy becomes independent of $q$, yielding $S_q = \log n$.

In our analysis, we use Renyi entropy as a multiscale diagnostic of the angular distribution of quasars on the sky. For each comoving distance $r$, we project all quasars within that volume onto the celestial sphere using the HEALPix pixelization scheme \citep{gorski05}, with a resolution parameter $N_{\mathrm{side}}$. This divides the sky into $N_{\mathrm{pix}} = 12 \times N_{\mathrm{side}}^2$ equal-area pixels.

Let $N_{\mathrm{eff}}$ be the number of populated pixels (i.e., pixels that contain at least one quasar) after masking and radial selection. For each pixel $i$, we compute the normalized angular density as
$f_i = \frac{n_i}{N}$, where $n_i$ is the number of quasars in $i^{\text{th}}$ pixel, and $N$ is the total number of quasars within the distance $r$. The Renyi entropy of order $q$ for the cumulative quasar distribution out to radius $r$ is then given by

\begin{equation}
S_q(r) = \frac{1}{1 - q} \log \sum_{i=1}^{N_{\mathrm{eff}}} f_i^q.
\label{eq:renyi_ent}
\end{equation}

By varying $q$, we modulate the sensitivity of the entropy measure: low $q$ values emphasize diffuse, underdense regions, while high $q$ values are more sensitive to dominant peaks or clustering. In an isotropic distribution, the entropies at different orders converge. Divergence among them signals anisotropy especially when it systematically varies with cosmic scale.

We limit our analysis to entropy orders  $q = 1, 2, 3, 4, 5$ as higher values of $q$ increasingly amplify the influence of rare, highly populated pixels effectively overemphasizing extreme clustering while suppressing sensitivity to the broader structure of the distribution. Very high $q$ values can become numerically unstable and highly sensitive to shot noise or outlier pixels. By focusing on low to moderate values of $q$, we strike a balance between capturing subtle variations in angular structure and maintaining statistical robustness. This range allows us to probe both diffuse features and localized clustering without being dominated by statistical noise or sampling artifacts.

This formulation enables us to capture scale-dependent, higher-order structural information in the angular distribution of quasars, and to compare it rigorously with randomized mock catalogs that assume statistical isotropy. Renyi entropy has been used earlier to test homogeneity in the galaxy distribution \citep{pandey21b, pandey21a}.

To explore how the Renyi entropies of different orders evolve with comoving scale, we plot $S_q(r)$ as a function of radial distance $r$ for $q = 1, 2, 3, 4, 5$ in \autoref{fig:combined_ent} across all three quasar samples. The left column displays the results for the real Gaia-unWISE quasar distribution, while the right column shows the corresponding behaviour for isotropic mock realizations. In each case, entropy increases with distance as expected, reflecting the accumulation of more uniform structure at larger scales. However, notable differences emerge in how the entropies of different orders converge. For the mock samples, the $S_q(r)$ curves for all $q$ converge rapidly with increasing $r$, indicating an angular distribution consistent with statistical isotropy. In contrast, the real data particularly for Sample~2 exhibits persistent separation between the entropy orders even at large $r$, signaling a degree of directional structure absent in the isotropic mocks. This divergence suggests that higher-order clustering features leave a measurable imprint in the quasar sky that standard two-point statistics may not capture.

To quantify this behaviour systematically, we introduce the normalized entropy dispersion $\Delta_{\mathrm{norm}}(r)$ and the corresponding significance ratio $\Psi(r)$ in the next subsection. These metrics will enable us to compress the full $S_q(r)$ curves into a single diagnostic that is sensitive to the relative spread of entropy across orders. 

\begin{figure*}[htbp!]
\centering
\includegraphics[width=14cm]{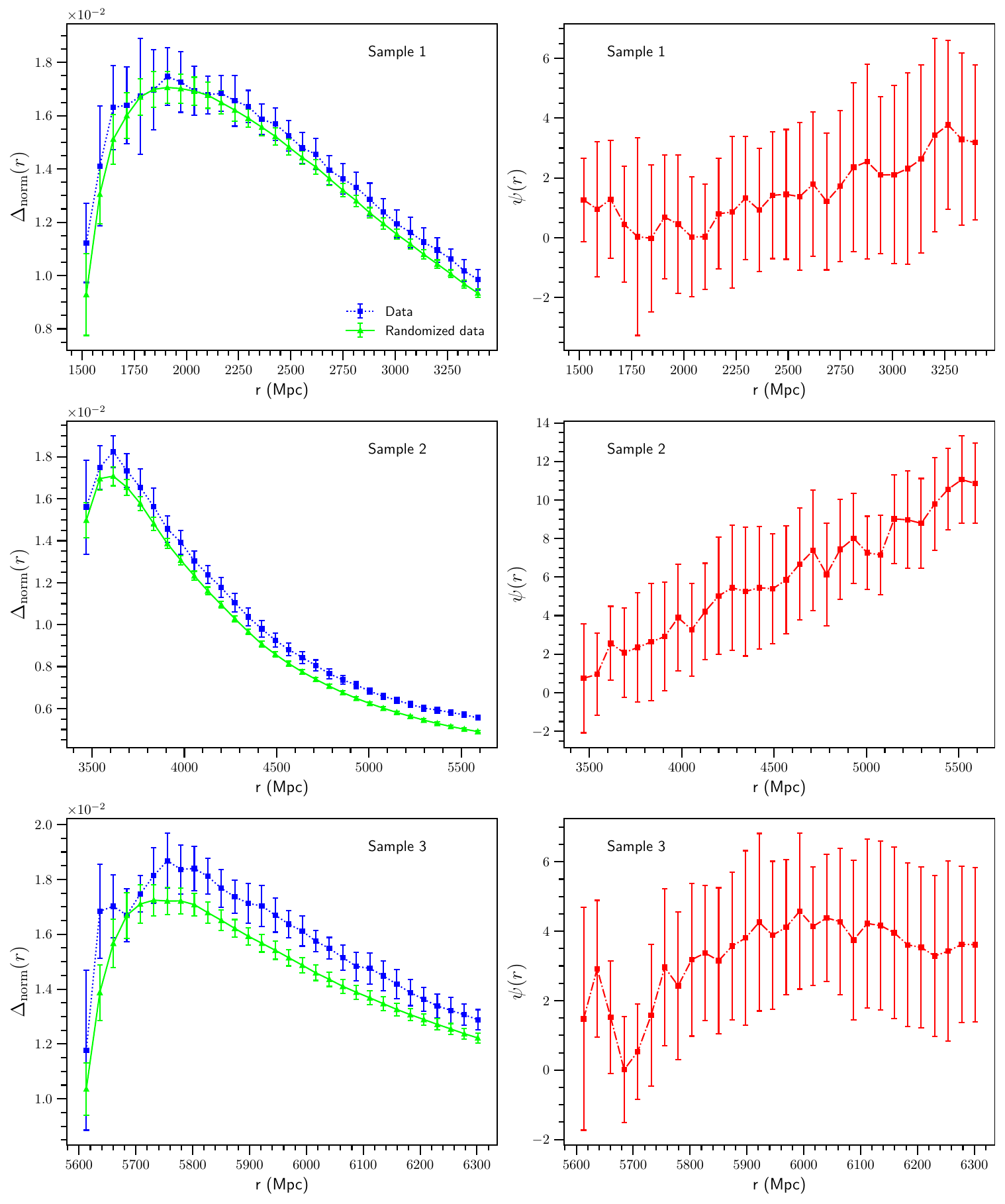}
\caption{This shows the impact of HEALPix resolution on anisotropy measures using Mask~1. Left panels show the mean normalized entropy dispersion for data and mocks as a function of radial distance. Right panels show the corresponding significance ratio $\Psi(r)$ with $1\sigma$ error bars. Top, middle, and bottom rows correspond to Samples~1, 2, and 3, respectively. The analysis uses $N_{\mathrm{side}} = 64$.}
\label{fig:64mask1}
\end{figure*}

\begin{figure*}[htbp!]
\centering
\includegraphics[width=14cm]{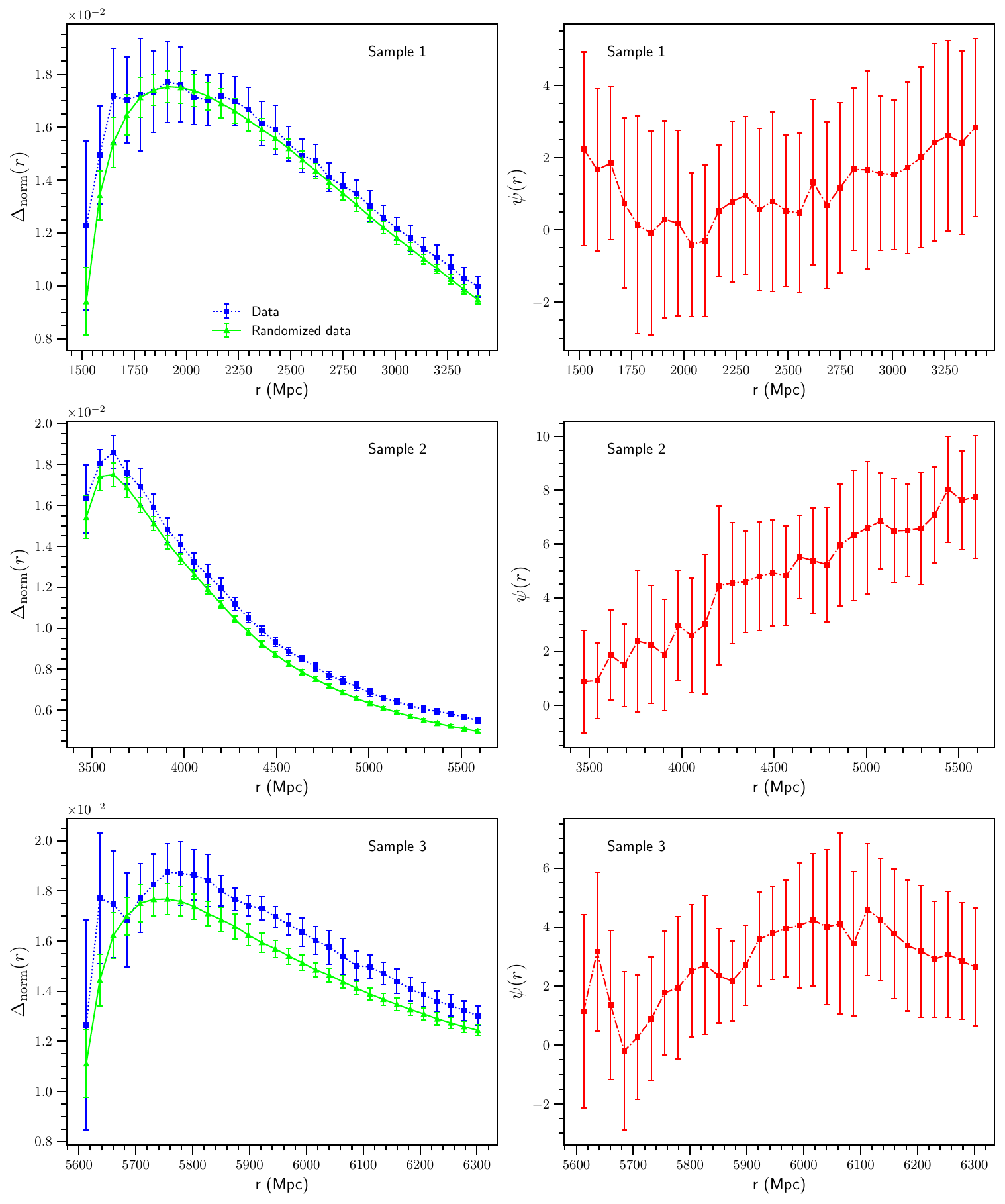}
\caption{Same as \autoref{fig:64mask1}, but using Mask~2 (galactic latitude cut plus circular mask). The plots illustrate that the main anisotropy signal, particularly in Sample~2, is preserved at higher angular resolution and under stricter masking.}
\label{fig:64mask2}
\end{figure*}

\subsection*{Normalized Entropy Dispersion}
To quantify scale-dependent deviations from isotropy in the angular distribution of quasars, we define the normalized entropy dispersion, denoted as $\Delta_{\mathrm{norm}}(r)$. This quantity measures the variation in Renyi entropy values across different entropy orders $q$ at a given comoving distance $r$. In a statistically isotropic distribution, all Renyi entropies regardless of $q$ should converge. Significant divergence among them signals directional structure in the quasar distribution.

Let $S_q(r)$ represent the Renyi entropy of order $q$ calculated from the angular distribution of quasars within radius $r$. The mean entropy across all $Q$ entropy orders is given by

\begin{equation}
S_{\mathrm{mean}}(r) = \frac{1}{Q} \sum_{q=1}^{Q} S_q(r),
\end{equation}

and the normalized entropy dispersion is

\begin{equation}
\Delta_{\mathrm{norm}}(r) = \frac{1}{S_{\mathrm{mean}}(r)} \sqrt{ \frac{1}{Q} \sum_{q=1}^{Q} \left( S_q(r) - S_{\mathrm{mean}}(r) \right)^2 }.
\label{eq:norm_entropy_disp}
\end{equation}

In this work, we consider five entropy orders ($Q = 5$), typically $q = 1, 2, 3, 4, 5$, enabling us to probe sensitivity to different aspects of the angular quasar distribution ranging from diffuse underdensities to strongly clustered regions. For each  cumulative comoving distance, we compute Renyi entropy across the same set of $q$ values and calculate $\Delta_{\mathrm{norm}}(r)$. The resulting \( \Delta_{\mathrm{norm}}(r) \) serves as a robust and scale-resolved indicator of anisotropy.

\begin{figure*}[htbp!]
\centering
\includegraphics[width=10cm]{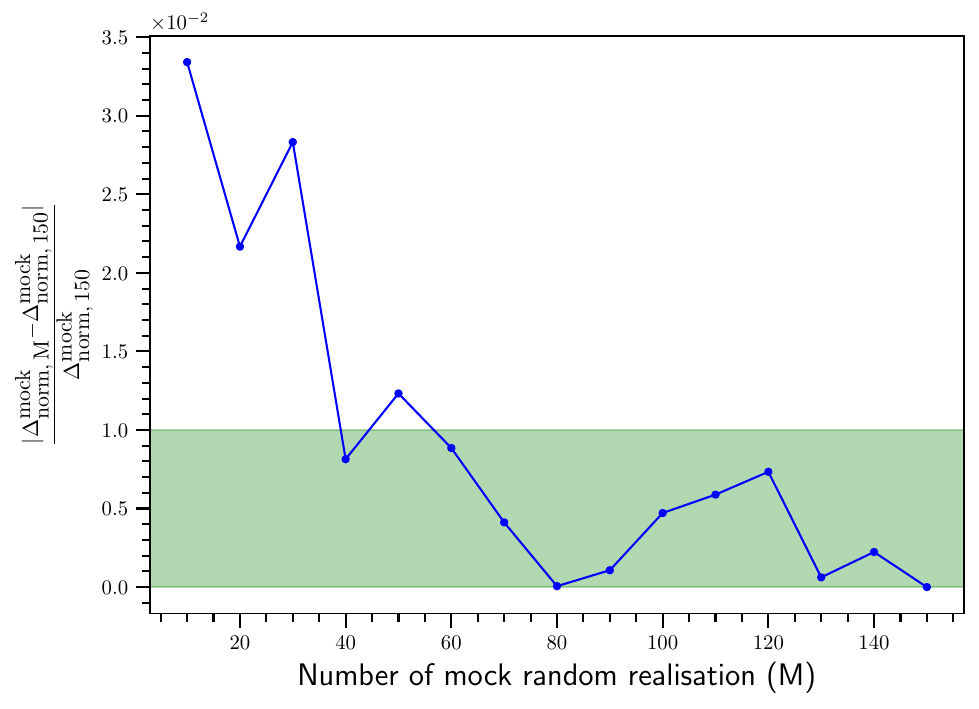}
\caption{This shows the convergence of the mean normalized entropy dispersion $\Delta_{\mathrm{norm}}^{\mathrm{mock}}(r)$ from mock realizations for Sample~2 at $r = 5000$ Mpc, using Mask~1. The plot shows the fractional deviation from the reference mean (computed using 150 realizations) as a function of the number of mocks M. The shaded band indicates the $1\%$ convergence threshold. The deviation falls below this level for M$\gtrsim 50$, demonstrating that the mock ensemble stabilizes beyond this point. Similar convergence behaviour is observed for other combinations of samples and masking schemes.}
\label{fig:converge}
\end{figure*}

\subsection*{Significance ratio and statistical comparison with mocks}
To interpret the normalized entropy dispersion in a statistically meaningful way, we compare it to the baseline expectation from isotropic realizations. Specifically, we generate 50 randomized mock quasar catalogues that preserve the radial selection function of the data but have angular coordinates drawn uniformly within the masked sky region. From these mocks, we compute the mean and standard deviation of the normalized entropy dispersion, denoted as $\Delta_{\mathrm{norm}}^{\mathrm{mock}}(r)$ and $\sigma_{\Delta^{\mathrm{mock}}_{\mathrm{norm}}}(r)$, respectively.

We also estimate the normalized entropy dispersion from the real quasar data, denoted $\Delta_{\mathrm{norm}}^{\mathrm{data}}(r)$, with uncertainties $\sigma_{\Delta^{\mathrm{data}}_{\mathrm{norm}}}(r)$ computed using 10 bootstrap resamplings within each redshift bin.

To assess how significantly the observed quasar distribution deviates from isotropy, we define a scale-dependent significance ratio, $\Psi(r)$, as

\begin{equation}
\Psi(r) = \frac{\Delta_{\mathrm{norm}}^{\mathrm{data}}(r) - \Delta_{\mathrm{norm}}^{\mathrm{mock}}(r)}{\sigma_{\Delta^{\mathrm{mock}}_{\mathrm{norm}}}(r)}.
\end{equation}

A value of $\Psi(r) = 0$ indicates perfect agreement between data and isotropic expectation, while $\Psi(r) = 1$ implies a one-sigma excess. Higher values suggest statistically significant anisotropy in the quasar angular distribution at scale $r$.

The uncertainty in $\Psi(r)$ is propagated using standard Gaussian error propagation

\begin{equation}
\sigma_{\Psi(r)} = \frac{\sqrt{ \sigma^2_{\Delta^{\mathrm{data}}_{\mathrm{norm}}}(r) + \sigma^2_{\Delta^{\mathrm{mock}}_{\mathrm{norm}}}(r) }}{ \sigma_{\Delta^{\mathrm{mock}}_{\mathrm{norm}}}(r) }.
\end{equation}

This framework allows us to track not only where entropy dispersion peaks, but also whether those deviations are statistically meaningful, thereby offering a rigorous lens into the scale-dependent isotropy of the Universe.


\section*{Validation and robustness tests}
To ensure the reliability and reproducibility of our entropy-based analysis, we conducted a series of validation tests designed to assess its sensitivity to key methodological choices. Specifically, we examine whether our results remain robust under changes in angular resolution and the number of isotropic mock realizations used to estimate statistical significance. 

\subsection*{Sensitivity to the HEALPix resolution parameter}
To assess the robustness of our findings with respect to angular resolution, we repeated our analysis using a higher HEALPix pixelization resolution, setting the resolution parameter to $N_{\mathrm{side}} = 64$. This finer resolution results in a much larger number of smaller pixels, enabling a more detailed probe of small-scale angular features in the quasar distribution.

We consider two masking strategies in this analysis. In \autoref{fig:64mask1}, we apply a simple galactic latitude cut, retaining quasars with $b \geq 40^\circ$ or $b \leq -60^\circ$, where $b$ denotes the Galactic latitude. Unlike the lower-resolution case ($N_{\mathrm{side}} = 8$), we do not implement additional completeness-based pixel selection since at this finer resolution most pixels already contain at least one quasar, making further sub-pixelization unnecessary.

\autoref{fig:64mask2} shows results obtained using a more conservative masking strategy. Here, we apply the same galactic latitude mask described above, along with an additional circular mask designed to exclude regions potentially affected by photometric systematics or stellar contamination, as detailed earlier in the preparation of Mask 2.

Across both masking schemes, we observe that the qualitative behaviour of the normalized entropy dispersion and the statistical trends of the significance ratio remain stable. Most notably, the prominent anisotropy signal in the intermediate redshift bin (Sample~2) persists at nearly the same level of statistical significance. This consistency reinforces the conclusion that the detected signal is not an artifact of pixel scale or angular binning.

An interesting trend emerges at the higher resolution of $N_{\mathrm{side}} = 64$. The normalized entropy dispersion $\Delta_{\mathrm{norm}}(r)$ exhibits a non-monotonic behaviour where it first increases and then decreases with radial distance. This contrasts with the monotonic decline observed at lower resolution. At finer angular scales, more detailed structures become visible. Initially, these enhance entropy dispersion by revealing small-scale clustering and voids, but beyond a certain distance, as the angular projection of structure saturates or dilutes, the dispersion decreases again. Despite this richer structure in the entropy statistics, the significance ratio $\Psi(r)$ which quantifies deviation from isotropy relative to randomized mocks remains robust across resolutions, since the mocks are subject to the same resolution and masking.

This validation confirms that our key findings, particularly the transitional epoch of anisotropy in Sample~2, are not dependent on a specific choice of HEALPix resolution. Our entropy-based approach captures meaningful large-scale anisotropies that persist under variations in angular resolution, pixel scale, and masking.

\subsection*{Convergence of mock entropy dispersion estimates}
To validate the robustness of our statistical baseline, we tested the convergence of the mean normalized entropy dispersion obtained from isotropic mock realizations. This step is essential to ensure that the ensemble average used in the significance ratio $\Psi(r)$ is stable and not affected by sampling noise due to an insufficient number of mocks.

We carried out a convergence test for Sample~2 and Mask~1 at a representative comoving radius of $r = 5000$\,Mpc, where the anisotropy signal is particularly strong  (\autoref{fig:dispersion1}). For each mock ensemble size M$= 10, 20, \ldots, 150$, we generated M independent randomized angular realizations of the quasar distribution, preserving the radial positions. For each value of M, we computed the mean normalized entropy dispersion $\Delta_{\mathrm{norm},\text{M}}^{\mathrm{mock}}(r)$ and compared it against the reference value $\Delta_{\mathrm{norm, 150}}^{\mathrm{mock}}(r)$ obtained using the full set of 150 mocks. \autoref{fig:converge} shows that the mock mean entropy dispersion converges rapidly with increasing $M$. We find that the fractional deviation from the reference mean (based on 150 mocks) drops below $1\%$ once the mock count exceeds approximately 50. Based on this convergence test, we adopted 100 mock realizations for the main analysis, which comfortably exceeds the threshold for statistical stability and ensures reliable estimation of the mock entropy dispersion across all radial scales. Repeating this test for other samples and masking schemes yields similar results, confirming that our significance calculations are not limited by the size of the mock ensemble.

This convergence test further reinforces the validity of our significance estimates and demonstrates that our results are not sensitive to the precise number of mocks used, provided the ensemble size is sufficiently large. The rapid convergence of the mock baseline also reflects the stability of the entropy-based measures across randomized angular distributions.

\section*{Acknowledgement}
AM thanks UGC, Government of India for support through a Junior Research Fellowship. BP acknowledges financial support from the SERB, DST, Government of India through the project CRG/2019/001110. BP would also like to acknowledge IUCAA, Pune for providing support through associateship programme and ICARD, Visva-Bharati.  

\subsection*{Data and Code Availability}

The Gaia-unWISE quasar catalogue used in this study is publicly available at \url{https://zenodo.org/records/8060755}. The codes used for this analysis are made available through a public GitHub repository:\url{https://github.com/amit280695/renyi-entropy-quasars} .

\end{document}